\documentstyle[12pt, a4]{article}
\begin{document}
\author{M. Ko\c{c}ak
\and Department of Engineering Physics, Faculty of Engineering,
\and University of Gaziantep, 27310 Gaziantep -T\"{u}rkiye}
\title{Bound State Solutions of Klein-Gordon Equation with the Kratzer Potential }
\date{}
\maketitle

The relativistic problem of spinless particle subject to a Kratzer
potential is analyzed. Bound state solutions for the
\textit{s}-wave are found by separating the Klein-Gordon equation
in two parts, unlike the similar works in the literature, which
provides one to see explicitly the relativistic contributions, if
any, to the solution in the non-relativistic limit.
\newline


\newline

The solution of Kratzer potential within the frame of
non-relativistic physics is well known in the literature, see for
instance \cite{gonul1}. However, due to the significance of exact
solutions for the relativistic equations in studying the systems
under the influence of strong potentials through the different
disciplines of physics, an increasing interest of the Klein-Gordon
(K-G) and Dirac equations has appeared in the last few  years  [
2-5 ]. By means of this interest the solution of Kratzer potential
is recently investigated \cite{wenchao} in the light of K-G
equation, but only for the consideration of equal scalar and
vector potentials leading to approximate energy solutions. In this
Letter, which is based on the recent discussion in \cite{gonul2},
we consider the general case where the scalar potential is unequal
to the vector potential, bearing in mind the existence of bound
states. The results obtained are compared to those in
\cite{wenchao} to clarify the importance of the present formalism
which reveals that the consideration of mixed equal potentials
such as \cite{wenchao} does not in usual reproduce the
relativistic effects. In fact, such calculations give solely an
idea about the appearance of K-G equations in the non-relativistic
border.

In the presence of vector and scalar potentials the
(1+1)-dimensional time-independent K-G equation for a spinless
particle of rest mass $m$  reads
\begin{equation}
-(\hbar c)^{2}\Psi''_{n}+(m
c^{2}+V_{S})^{2}\Psi_{n}=(E_{n}-V_{V})^{2}\Psi_{n}~~,
~~~n=0,1,2,...
\end{equation}
where $E$ is the relativistic energy of the particle, $c$ is the
velocity of the light and $\hbar$ is the Planck constant. The
vector and scalar potentials are given by $V_{V}$ and $V_{S}$,
respectively. The subscripts for the terms of potential indicate
their properties under a Lorentz transformation: $V$ for the time
component of the 2-vector potential ad $S$ for the scalar term.

In the non-relativistic approximation (potential energies small
compared to $mc^{2}$  and $E \cong mc^{2}$) Eq. (1) becomes
\begin{equation}
-\frac{\hbar^{2}}{2m}\Psi''_{n}+(V_{S}+V_{V})\Psi_{n}\approx(E_{n}-m
c^{2})\Psi_{n}~.
\end{equation}
Eq. (2) show us that $\Psi$ obeys the Schr\"{o}dinger equation
with binding energy equal to $E-m c^{2}$ , and without
distinguishing the contributions of vector and scalar potentials.

From \cite{gonul2}, the full relativistic wave function in (1) may
be expressed as $\Psi=\chi\phi$  where $\chi$ represents the
behavior of the wave function in the non-relativistic region and
$\phi$ is the modification function due to the relativistic
effects. This consideration opens a gate to transform Eq. (1) into
a couple of equation ($\hbar=c=1$ ),
\begin{equation}
\frac{\chi''_{n}}{\chi_{n}}=2(m V_{S}+E_{n}
V_{V})-\varepsilon_{n}~,
\end{equation}
\begin{equation}
\frac{\phi''_{n}}{\phi_{n}}+2\frac{\chi'}{\chi}\frac{\phi'_{n}}{\phi_{n}}=(V_{S}^{2}-V_{V}^{2})-\Delta\varepsilon_{n}~,
\end{equation}
in which, considering (2), Eq. (3) corresponds to the form of K-G
equation in the non-relativistic limit while (4) denotes the
corrections to the non-relativistic solutions. In addition,
$\varepsilon$ and $\Delta\varepsilon$ represent the binding energy
within the non-relativistic domain and the modification term
because of the relativistic consideration, respectively, yielding
$E^{2}-m^{2}=\varepsilon+\Delta\varepsilon$. It is noticeable that
relativistic contributions disappear in case of $V_{V}=\pm V_{S}$
as in \cite{wenchao}. Thus Eq (3) reduces to a free particle
problem if $V_{V}=-V_{S}$ because $E \approx m$ in the limit where
this equation valid while it reproduces Schr\"{o}dinger like
non-relativistic solutions for the case $V_{S}=V_{V}$, which
overall justify the reliability of the formalism when the ongoing
discussions considered in the literature, e.g. [2,4].

For practical calculations, Eqs.  (3) and (4) are expressed by the
Riccati equation
\begin{equation}
W^{2}_{n}-\frac{W'_{n}}{\sqrt{2m}}=2(mV_{S}+E_{n}V_{V})-\varepsilon_{n},
~~~~W_{n}=-\frac{1}{\sqrt{2m}}\frac{\chi'_{n}}{\chi_{n}}~,
\end{equation}
\begin{equation}
\Delta W^{2}_{n}-\frac{\Delta W'_{n}}{\sqrt{2m}}+2W_{n}\Delta
W_{n}= (V_{S}^{2}-V_{V}^{2})-\Delta\varepsilon_{n}, ~~~~ \Delta
W_{n}= -\frac{1}{\sqrt{2m}}\frac{\phi'_{n}}{\phi_{n}}~.
\end{equation}
Note that if the whole potential
$2mV_{S}+2E_{n}V_{V}+V_{S}^{2}-V_{V}^{2}$ is an exactly solvable
then the above equations turn out to be a simple form within the
framework of the usual supersymmetric quantum theory
\cite{cooper}, where $W^{SUSY}_{n}=W_{n}+\Delta W_{n}$. The reader
is referred to \cite{gonul1} to see such solution of the Kratzer
potential in the arbitrary dimensions. However , if Eq. (6) has no
analytical solution one cannot use $W^{SUSY}_{n}$ concept in
dealing with such problems. To overcome this drawback of the
formalism, an elegant reliable technique leading to approximate
solutions of (6) has been recently introduced in Ref.
\cite{gonul3} for any state of interest.

Now, let us focus on the scalar and vector potentials in the form
\begin{equation}
V_{S}=\frac{A_{1}}{r^{2}}-\frac{B_{1}}{r} , ~~~~
V_{V}=\frac{A_{2}}{r^{2}}-\frac{B_{2}}{r} ,
\end{equation}
which, in the light of Eq. (5), restricts us to define
\begin{equation}
W_{n=0}=-\frac{c+1}{r}+\frac{k}{2(c+1)} , ~~ c>0, ~k>0
\end{equation}
where $k=2mB_{1}+2EB_{2}$ and
\begin{equation}
c=\frac{(A_{1}B_{1}-A_{2}B_{2})}{\sqrt{A^{2}_{1}-A^{2}_{2}}}
\end{equation}
or
\begin{equation}
c(c+1)=2mA_{1}+2EA_{2}
\end{equation}
thus,
\begin{equation}
c=-\frac{1}{2}+\sqrt{\frac{1}{4}+2(mA_{1}+EA_{2})}.
\end{equation}
Although we have two definition of $c$, (9) and (11), we will use
only the physically acceptable one which is (11). Because (9),
which has no physical meaning, doesn't reproduce physically
acceptable  $c$-values when compared to the works of Castro
\cite{castro} in case $A_{1}=A_{2}=0$. Thus, the corresponding
full non-relativistic energy spectrum and unnormalized wave
function in the ground state are in the form of
\begin{equation}
\varepsilon_{n}=-\frac{k^{2}}{4(n+c+1)^{2}} , ~~
\chi_{n=0}=r^{c+1}e^{-\frac{kr}{2(c+1)}} ,
\end{equation}
which are in agreement with those in \cite{gonul1} that was
performed in the non-relativistic frame.

With the consideration of (6), we set $\Delta W$ as
\begin{equation}
\Delta W=-\frac{a}{r^{2}}~, ~~a>0
\end{equation}
where $a=\sqrt{A^{2}_{1}-A^{2}_{2}}$. The procedure until here
shows us that $A_{1}>A_{2}$, $|B_{1}|<|B_{2}|$, and thus the
requirement for bound states such that $V_{S}>V_{V}$ and
$E^{2}-m^{2}=\varepsilon+\Delta \varepsilon<0$, subsequently $m>E$
are satisfied. For the case $V_{V}=\pm V_{S}$ , $a$ vanishes,
consequently $\Delta W\rightarrow 0$, together with $\Delta
V\rightarrow 0$. From equation (6) and (11), in case $a>0$, the
relativistic contributions to the non-relativistic solutions are
\begin{equation}
\Delta\varepsilon_{n}=0 ,~~~ \phi_{n=0}=e^{-\sqrt{2m}\int{\Delta
W_{n=0}dz}}=e^{-\frac{a}{r}} .
\end{equation}
Hence, the full solutions corresponding the total potential
$2mV_{S}+2E_{n}V_{V}+V^{2}_{S}-V^{2}_{V}$ are
\begin{equation}
E^{2}_{n}-m^{2}=\varepsilon_{n}+\Delta\varepsilon_{n}=-\frac{k^{2}}{4(n+c+1)^{2}}=
-\frac{4(mB_{1}+E_{n}B_{2})^{2}}{\left[2n+1+\sqrt{1+8(mA_{1}+E_{n}A_{2})}\right]^{2}}<0,
\end{equation}
\begin{equation}
\psi_{n=0}=\chi_{n=0}\phi_{n=0}=r^{c+1}\exp{\left(-\frac{a}{r}-\frac{kr}{2(c+1)}\right)}.
\end{equation}
Though the energy correction is zero in this specifically chosen
example, however this is not the case in general for other
problems \cite{gonul2}. It is stressed at this point that one can
directly solve infact the K-G equation, without use of a
separation procedure as in the present scheme, employing the total
form of $W^{SUSY}_{n=0}$ above in a Riccati equation similar to
(5) in connection with the whole potential. However, such a
treatment is not so practical due to the screening of the
relativistic contributions in the calculation results.

To test the reliability of  (15) let $A_{1}=A_{2}=0$, then
\begin{equation}
E_{n}=m\frac{-\frac{B_{1}B_{2}}{(n+1)^{2}}\pm\sqrt{1-\frac{B^{2}_{1}-B^{2}_{2}}{(n+1)^{2}}}}
{1+\frac{B^{2}_{2}}{(n+1)^{2}}}~,
\end{equation}
that overlaps with result in \cite{castro}.

Since we know the solution of the problem, we will start to
analyze some special cases.

\textbf{(1)} In the case of a pure scalar potential, $V_{V}=0$,
where we have $A_{2}=B_{2}=0$ and
$V=V_{S}=\frac{A_{1}}{r^{2}}-\frac{B_{1}}{r}>0$,
\begin{equation}
E_{n}=\pm
m\sqrt{1-\frac{4B^{2}_{1}}{\left[2n+1+\sqrt{1+8mA_{1}}\right]^{2}}}~,
\end{equation}
so that the bound state energy levels for particles and
antiparticles are symmetric about $E_{n=0}$. If $A_{1}=0$ then Eq.
(18) reduces to the solution in \cite{castro}.

\textbf{(2)} In the pure vector potential case, $V_{S}=0$, we have
only a potential term
$V=V_{V}=\frac{A_{2}}{r^{2}}-\frac{B_{2}}{r}$ ; $A_{1}=B_{1}=0$.
Then we have bound state energy
\begin{equation}
E^{2}_{n}-m^{2}=-\frac{4E^{2}_{n}B^{2}_{2}}{\left[2n+1+\sqrt{1+8E_{n}A_{2}}\right]^{2}}~,
\end{equation}
the expression of $E_{n}$ is so complicated in order to to observe
its physical meaning. Because of this we use power series to have
an approximate energy solution
\begin{equation}
E_{n}=m\left[1-\frac{B^{2}_{2}}{2(n+1+2mA_{2})^{2}}\right].
\end{equation}
In this circumstances, the energy spectrum consists of energy
levels either for particle $V_{V}>0$ or for antiparticles
$V_{V}<0$. To compare with the study in Ref. \cite{castro} we
choose $A_{2}=0$ then (16) will have same bound state energy
\begin{equation}
E_{n}=\pm\frac{m}{\sqrt{1+\frac{B^{2}_{2}}{(n+1)^{2}}}} .
\end{equation}

\textbf{(3)} For $V_{V}=\pm V_{S}$, it is note that relativistic
contributions disappear in this case. So we are dealing only with
the non-relativistic limit solutions. There are two conditions
which first one is \textbf{(i)} $V_{V}=V_{S}$, it requires the
equality of parameters of both potentials, $A_{2}=A_{1}$ and
$B_{2}=B_{1}$ , $c(c+1)=2mA_{1}+2EA_{2}=2(m+E)A_{1}$ with
$k=2(m+E)B_{1}$. In the light of this point and together with Eq.
(15) one obtains
\begin{equation}
E^{2}_{n}-m^{2}=-\frac{4(m+E_{n})^{2}B^{2}_{1}}{\left[2n+1+\sqrt{1+8(m+E_{n})A_{1}}\right]^{2}}~,
\end{equation}
to avoid from a complicated expression for $E_{n}$~, we set
$\sqrt{m-E_{n}}=\alpha$ and then expand (22) as a power series of
$\alpha$. Leaving out the $\alpha^{2}$ and higher terms because of
their negligible small values as compared to $\alpha$, we find
$\alpha=\frac{2\sqrt{2m}B_{1}}{\left[2n+1+\sqrt{1+16mA_{1}}\right]}$~,
and then from $m-E_{n}=\alpha^{2}$
\begin{equation}
E_{n}=m-\frac{8mB^{2}_{1}}{\left[2n+1+\sqrt{1+16mA_{1}}\right]^{2}}.
\end{equation}
This result agrees with the circumstance in Ref. \cite{wenchao}.
The second test is done by Castro \cite{castro}; when we assume
$A_{1}=0$ in (22) we get
\begin{equation}
E_{n}=m\left[\frac{(n+1)^{2}-B^{2}_{1}}{(n+1)^{2}+B^{2}_{1}}\right],
\end{equation}
that is same with the case in \cite{castro}. Energy levels
obtained in (23) and (24) correspond to bound states of particles.
In this case there are no energy levels for antiparticles. The
second condition for the third case is  \textbf{(ii)}
$V_{V}=-V_{S}$ where $A_{2}=-A_{1}$ and $B_{2}=-B_{1}$ then the
energy spectrum
\begin{equation}
E^{2}_{n}-m^{2}=-\frac{4(m-E_{n})^{2}B^{2}_{1}}{\left[2n+1+\sqrt{1+8(m-E_{n})A_{1}}\right]^{2}}~,
\end{equation}
as in the previous case we are using power series expansion about
the $\sqrt{m-E_{n}}=\alpha$ one obtains
\begin{equation}
E_{n}=-m\left(\frac{2(n+1)^{2}}{B^{2}_{1}}-1\right),
\end{equation}
which for $A_{1}=0$ Eq. (25) reduces to Castro's \cite{castro}
related situation
\begin{equation}
E_{n}=-m\left[\frac{(n+1)^{2}-B^{2}_{1}}{(n+1)^{2}+B^{2}_{1}}\right]
\end{equation}
in contrast to previous case , now  energy levels in (26) and (27)
correspond to antiparticles and so there is no any energy spectrum
for the particles.

The idea suggested in this Letter would be used to explore a great
number of relativistic systems and can be also extended to the
case of the Dirac equation that is now under consideration.

\end{document}